\documentclass[12pt]{article}
\begin{document} 
{\Huge An Algorithm to Simplify Tensor Expressions}

\

\begin{center}
R. Portugal\footnote{ On leave from Centro Brasileiro de Pesquisas
F\'{\i}sicas, Rua Dr. Xavier Sigaud 150, Urca, Rio de Janeiro, Brazil. CEP
22290-180. Email: portugal@cat.cbpf.br} 

{\small Department of Applied Mathematics, University of Waterloo,
Waterloo, Ontario, N2L 3G1 - Canada.}
\end{center}

\  

\

\begin{center}
Abstract
\end{center}

\

           The problem of simplifying tensor expressions is
addressed in two parts. The first part presents an  algorithm designed
to put  tensor expressions into a canonical form, taking into account
the symmetries with respect to index permutations and the renaming of
dummy indices. The tensor indices are split into classes and a natural
place for them is defined. The canonical form is the closest
configuration to the natural configuration. In the second part, the
Gr\"obner basis method  is used to simplify tensor 
expressions which obey the linear
identities that come from cyclic symmetries (or more general tensor
identities, including non-linear identities).  The algorithm is
suitable for implementation in general purpose computer algebra
systems. Some timings of an experimental implementation  over the
Riemann package are shown.

\

\

\section{Introduction}

Recently, some attempts to describe algorithms for simplifying tensor
expressions have appeared in the 
literature.\cite{Fulling}\cite{Kryukov}\cite{Dresse} 
The tensor symmetry properties and the presence of
dummy indices make the problem complex. Fulling et al.\cite{Fulling}
described an algorithm to enumerate the independent
monomials built from 
the Riemann tensor and its covariant derivative. They presented
explicit tables for  monomials of order 12 or less in derivatives
of the metric. Although explicit 
bases are presented for monomial of order 8 or less, they have 
not described neither a systematic algorithm to build the
basis nor a method to simplify generic tensor 
expressions built
using the Riemann tensor. Using the
same algorithm, Wybourne and Meller\cite{Wybourne} enumerated the order-14
invariants. On the other hand,  
Ilyin and Kryukov\cite{Kryukov} did present an algorithm to simplify
general tensor expressions based on the group algebra of
the permutation group.  Though the method is elegant from the
mathematical point of view, it is inefficient. Quoting 
the authors: ``hardware development is very fast
now, and it will be possible to solve problems with 11
indices with the help of our program''. 
Dresse presented in the second part of
his PhD dissertation\cite{Dresse} a new algorithm to simplify tensor
expression based on the backtrack algorithms for combinatorial
objects.\cite{Butler} Most of his effort has been directed to 
solve the ``dummy index problem'' (described bellow).

In this work, the problem is addressed in two parts. 
In the first, the tensor
expressions are put into a canonical form taking into account the
symmetries with respect to index permutations and renaming of dummy
indices. In the second part, the Gr\"obner basis method (see Geddes
at al.\cite{Geddes} and refs. therein) is used to address the
linear identities that come from cyclic symmetries (or more general
tensor identities). The algorithm is suitable for implementation in
general purpose computer algebra systems, since many functionalities
of these systems  (besides the Gr\"obner basis implementation) are
required.

It is known that the names of the dummy indices have no intrinsic
meaning, and the presence of two or more of them in a tensor
expression creates symmetries with respect to renaming. This leads to
algorithms of complexity 
${\rm O}(n{\rm !})$
where 
$n$ 
is the number of dummy indices. Dummy indices difficult the
definition of tensor rules and the use of pattern matching. One way to
solve the dummy index problem is to rename them in terms of the index
positions and the name of the tensors (see step 9 of the algorithm).
However, the tensor symmetries may change the index positions,
invalidating the process. This kind of renaming method is invariant if
there is a prescribed method to put the indices into a canonical
position. 

In section 2, an algorithm to put tensor expressions into a canonical
form is described. The canonical position of the indices is based on
definitions 3 to 6. Some parts of the definitions of this section
involve conventions that can be changed without losing the
``canonicalization'' character of the algorithm. The ideal format for
the canonical tensor is 

\begin{center}
 $T\,^{F^{+}}\,^{S_{II}}\,^{A^{+}}\,^{B^{+}}\,^{C^{+}}\,^{S_{I}}\,
{_{C^{-}}}\,{_{B^{-}}}\,{_{A^{-}}}\,{_{F^{-}}}$
\end{center}

\noindent where 
$F^{+}$, 
${S_{{\it II}}}$, 
$A^{+}$, 
$B^{+}$, 
$C^{+}$, 
${S_{I}}$, 
$C^{-}$, 
$B^{-}$, 
$A^{-}$
and 
$F^{-}$
represent sequences of indices whose meanings are defined in step 7
of the algorithm. The indices of class 
${S_{I}}$
can have contravariant or covariant character. 
They are placed in the midst
of the classes that have the character fixed.
This ideal format is only achieved in special cases,
as when the tensor 
$T$
is totally symmetric or antisymmetric. The canonical position of the
indices is the one closest  to this format, where the notion of 
``closest'' is precisely defined.

Product of equal tensors is addressed by reduction to one tensor 
of rank equal to the sum of the rank of the factors. 
This reduction process is used in
other parts of the algorithm in order to simplify the
implementation.

In section 3, some timings of an experimental implementation of the
algorithm over the Riemann package\cite{Portugal} are presented. 
Polynomials built from the Riemann tensor are challenges for any
algorithm to simplify tensor expressions.  I believe that special
techniques can improve the timings for the kind of symmetries of the
Riemann tensor, but none has been implemented for the demonstration of
this section. The implementation in the Maple system\cite{Maple}
can easely handle products of three Riemann tensors.

In section 4, the problem of the simplification of tensor expressions
under the presence of side tensor identities, which come generally
from cyclic symmetries, is addressed. The Gr\"obner basis
implementation of the Maple system is used to accomplish the full
simplification.

\section{The algorithm}

Consider the definition of the normal and canonical functions given in
Geddes et al.\cite{Geddes} Let 
$\mathcal{A}$
be a set of algebraic 
expressions which admits a canonical function. Consider
the operations of multiplication, addition and contraction of tensors
as defined in the tensor algebra.\cite{Rund}\cite{Eisenhart}\cite{Thomas} 
If a coordinate system has been selected, the
tensor algebra can be performed through the tensor components. In this
work, a tensor expression is any expression written in terms of
non-assigned tensor components obeying the rules of the tensor algebra
whose coefficient factors are members of the set 
$\mathcal{A}$. Consider Riemannian spaces, in which exists a
fundamental metric tensor which establish a relation between the
covariant and contravariant tensor indices.

\subsection{Definitions}

\noindent \textbf{Hypothesis 1}: Tensors do not obey any side
tensor identities except symmetries with respect to index
permutations or symmetries with respect to renaming or the
inversion of character of dummy indices.

``Symmetries with respect to index permutations'' means that the
tensors obey one or more equations of the kind

\begin{center}
$T\,^{{i_{1}}}\,^{\cdots}\,^{{i_{n}}}=\epsilon \,T\,^{\pi (
{i_{1}}\,\cdots\,{i_{n}})}$, 
\end{center}

\noindent where 
$\epsilon=$
1 or
$\epsilon=-$
1 and 
$\pi ({i_{1}}\,\cdots\,{i_{n}})$
 is any permutation of  
${i_{1}}\,\cdots\,{i_{n}}$.

\

\noindent \textbf{Definition 1}: \textit{Induced symmetry} of a sub-set of
indices of a tensor. 

The \textit{induced symmetries} of a sub-set of indices of a tensor
are the symmetries that the sub-set inherits from the symmetries of
the whole set of indices. Pairs of dummy indices are treated as
independent free indices, hence not permutable.

   For example, the \textit{induced symmetry} of the first two indices
of  the Riemann tensor is the skew symmetry. The second and third
indices have no  \textit{induced symmetry} regardless any 
contraction between the first and
fourth indices.

\

\noindent \textbf{Definition 2}:  \textit{Equivalent} index configurations.

Two index configurations\footnote{ The index configuration is the list
of indices of the tensor, taking into account
the character of each index.} of a tensor 
$T$
 extracted from a tensor product are said to be \textit{equivalent}
if one configuration can be put into the other by the use of any of
the following properties:

1. Character inversion of the dummy indices,

2. Renaming the dummy indices,

3. Index permutation allowed by the symmetries of the tensors of the
tensor product.

\

\noindent \textbf{Definition 3}:  Suppose the tensor 
$T$
has 
$n$
indices and let 
(${\lambda _{1}}$, $\cdots$, ${\lambda _{p}}$) 
be a partition of 
$n$
 where 
$p$
 is a positive integer less or equal than 
$n$. The indices of 
$T$
 can be grouped in disjoint classes 
${C_{1}}$,
 $\cdots$, 
${C_{p}}$
where a generic class  
${C_{i}}$
has 
${\lambda _{i}}$
indices.      The indices of each class can be substituted with
numbers in such way that the indices of class 
${C_{i}}$
run from 
$(\sum _{j=1}^{i - 1}\,{\lambda _{j}}) + 1$
 to 
$\sum _{j=1}^{i}\,{\lambda _{j}}$. 
Consider all index configurations of the tensor 
$T$
and let the indices be substituted with the corresponding numbers.
The configurations are in one-to-one correspondence with the elements
of the symmetric group 
${S_{n}}$.\cite{Littlewood}\cite{Waerden} The following criteria establish
an order of decreasing configurations with respect to classes  
${C_{1}}$
to  
${C_{p}}$:

\textbf{a.}  Smaller value of the position of the first index of class
${C_{1}}$
in the tensor 
$T$. If the positions are equal, consider the position of the next index
of class  
${C_{1}}$. If the positions of all indices of class  
${C_{1}}$
are equal, consider the positions of the indices of the next classes
until  
${C_{p}}$.

\textbf{b}.  Smaller value of the first index member of class 
${C_{1}}$
that appears in the tensor 
$T$.  If the first indices of class 
${C_{1}}$
that appear in all configurations are equal, consider the next index
member of class 
${C_{1}}$. If the indices of class 
${C_{1}}$
appear in the same order, consider the order of the indices of the
next classes until 
${C_{p}}$. 

 Definition 3a alone compares the position of the classes, while
definition 3b alone compares the order of the indices. Given a set of
\textit{equivalent} configurations of a tensor \textit{T}, definition
3 allows one to select the smallest configuration of the set with
respect to a given partition of the number of indices. The smallest
configuration is unique. If definition 3b is not applied, or if it is
applied for some but not all classes, instead of having one smallest
configuration, one may have a sub-set of smallest configurations. 

\

\noindent \textbf{Definition 4}: \textit{Character normal configurations}.

Let 
$C^{+}$
and 
$C^{-}$
be the classes of the contravariant and covariant indices
respectively of a tensor 
$T$
extracted from a tensor product. Consider the set of  all
\textit{equivalent} index configurations of \textit{T}.  The
\textit{character normal configurations} consist of the sub-set of
smallest index configurations according to definition 3a with respect
to classes 
$C^{+}$
and 
$C^{-}$. 

\

\noindent \textbf{Definition 5}: \textit{Index normal  configurations}.

Consider the definition of group I and II given in step 3 and the
definition of classes 
$F^{+}$, 
${S_{{\it II}}}$, 
$A^{+}$, 
$B^{+}$, 
$C^{+}$, 
${S_{I}}$,\footnote{ 
${S_{I}}$
 is a collection of sub-classes, which have the same status of the
other classes in this definition. The order of the sub-classes is
described in step 7c.} 
$C^{-}$, 
$B^{-}$, 
$A^{-}$
and 
$F^{-}$
given in step 7a.  The present criteria applies when a tensor
(\textit{T})  of group I is extracted from a tensor product. Consider
the set of  all \textit{equivalent} index configurations of
\textit{T}. Let the indices be relabelled as described in step 7c. The
 \textit{index normal configurations} of the tensor \textit{T} consist
of the sub-set of smallest configurations that are \textit{character
normal configurations}, and satisfy definition 3a for classes  
$F^{+}$, 
${S_{{\it II}}}$, 
$A^{+}$, 
$B^{+}$, 
$C^{+}$, 
${S_{I}}$, 
$C^{-}$, 
$B^{-}$, 
$A^{-}$
and 
$F^{-}$
and definition 3b  for classes 
$F^{+}$, 
${S_{{\it II}}}$, 
$B^{+}$, 
$C^{+}$,
${S_{I}\,^{0}}$, 
$C^{-}$,  
$B^{-}$
and  
$F^{-}$. 

\

\noindent \textbf{Definition 6}: \textit{Index canonical configuration}.

The  \textit{index canonical configuration} is the only element of the
set of  \textit{index normal configurations} which  fully satisfies
definition 3 with respect to classes  
$F^{+}$
, 
${S_{{\it II}}}$
, 
$A^{+}$
, 
$B^{+}$
, 
$C^{+}$
, 
${S_{I}}$
, 
$C^{-}$
, 
$B^{-}$
 , 
$A^{-}$
and 
$F^{-}$
in this order.  

\

The order of the indices of classes 
$A^{+}$
and  
$A^{-}$
and the order of the indices of the sub-classes of 
${S_{I}}$ (not  including ${S_{I}\,^{0}}$)
are not considered in the definition of  the  \textit{index normal
configurations}. In fact, if classes 
$A$
and 
${S_{I}}$
have less than two elements each,  
the set of  \textit{index normal
configurations} has one element, 
which is the \textit{index canonical configuration}.

An algorithm to simplify tensor expressions must recognize
when an expression vanishes. The following three lemmas
guarantee that the algorithm presented here can recognize
null tensor products.

\

\noindent \textbf{Lemma 1}: Let $\mathcal{P}$ be a product of 
non-null tensors and suppose that no factor 
vanishes. Let \textit{A} be
be a factor and define the tensor \textit{S} as

\[ S={\frac {\mathcal{P}}{A}}. 
\]

\noindent Suppose that \textit{A} and \textit{S} 
share \textit{n} 
contracted indices. Let \textit{s} be
the rank of \textit{S}, and 
\textit{a} be \textit{n} plus the number
of free indices of \textit{A}. 

    The product $\mathcal{P}$ is zero if and only if 
there exists a factor 
\textit{A} having the symmetry

\begin{equation}
A\,^{{i_{1}}\,\cdots\,{i_{a}}}= - A\,^{\pi ({i_{1}}
\,{\cdots}\,{i_{a}})}
\end{equation}

\noindent where the permutation $\pi$ acts only on the 
indices contracted with the 
indices of \textit{S}, 
and  $\mathcal{P}$ is invariant under the application
of the permutation $\pi$ on the corresponding indices of
\textit{S}, that is

\begin{equation}
\mathcal{P} = 
A\,^{{i_{1}}\,\cdots\,{i_{a}}}\,S\,{_{\pi ({j_{1}}\,\cdots
\,{j_{s}})}},
\end{equation}

\noindent where \textit{n}
\textit{j}'s are equal to \textit{n} 
\textit{i}'s and the permutation $\pi$ acts on names, not on positions.\footnote{The 
character of the indices
of \textit{A} and \textit{S} need not be contravariant and
covariant respectively. The only restriction is that the character 
of the dummy indices are opposite.} 
\textit{A} may have indices contracted internally which have
not been represented in (1) and (2). Symmetry (1)
takes account of permutations and character inversions
of the dummy indices within \textit{A}.

\noindent  \textbf{Proof:}  
($\Rightarrow$)(by \textit{reductio ad absurdum}) 
There are two cases: 

1. If $\mathcal{P}$ is not invariant 
for any factor \textit{A} that has the symmetry (1),
then

\[
\mathcal{P} \neq 
A\,^{{i_{1}}\,\cdots\,{i_{a}}}\,S\,{_{\pi ({j_{1}}\,\cdots
\,{j_{s}})}}.
\]

\noindent Using (1) and renaming the dummy indices, 
it follows that 
$\mathcal{P} \neq - \mathcal{P}$,
therefore 
$\mathcal{P} \neq 0$.

2. If no factor admits a symmetry $\pi$ of the form (1),
from the supposition that no factor of $\mathcal{P}$ 
vanishes, it follows that 
$\mathcal{P} \neq 0$.

\

So far only products of non-vanishing factors have
been considered. What are the conditions that 
cancel a single generic tensor with some or all
indices contracted? The answer can be obtained from
lemma 1. Suppose that \textit{T} is a tensor  of rank
$m$ with $2n$ indices contracted. This tensor can
be written as 

\begin{equation}
g\,{_{{i_{1}}{j_{1}}}}\,\cdots\,g\,{_{{i_{n}{j_{n}}}}
}\,T\,^{\sigma ({i_{1}}}\,^{{j_{1}}}\,^{\cdots
}\,^{{i_{n}}}\,^{j_{n}}\,^{f_{1}}\,^{\cdots}\,
^{f_{m-2n})},
\end{equation}

\noindent where $\sigma$ is the permutation of 
${{i_{1}}}\,{{j_{1}}}\,\cdots\,{{i_{n}}}\,{{j_{n}}}
\,{f_{1}}\,\cdots\,{f_{m-2n}}$
that specifies the actual order of the indices of \textit{T},
and $g\,{_{i}}\,{_{j}}$ is the metric supposedly symmetric. 
The free indices are ${f_{1}}\,\cdots\,{f_{m-2n}}$.
For this case, the
tensor \textit{S} of lemma 1 is

\[
S\,{_{{i_{1}}}}\,{_{{j_{1}}}}\,\cdots\,{_{{i_{n}}}}\,{_{{j
_{n}}}} = g\,{_{{i_{1}}{j_{1}}}}\,\cdots\,g\,{_{
{i_{n}{j_{n}}}}},
\]

\noindent and is symmetric under the interchange of 
$i\,{_{p}} j\,{_{p}}$ into
$j\,{_{p}} i\,{_{p}}$ for $p \leq n$,
and is totally symmetric under
the pair interchange of 
$i\,{_{p}} j\,{_{p}}$
into
$i\,{_{q}} j\,{_{q}}$
for $p,q \leq n$. Tensor \textit{S} does not vanish (again
from lemma 1).

  If the second factor of (3) does not vanish, 
there are two cases to consider regarding lemma 1. In the
first case, 
tensor \textit{T} has antisymmetry (1) while
$S$ is symmetric under the same permutation 
acting on the corresponding indices independently
of \textit{S} being multiplied by \textit{T}. In the second
case,  
the indices of \textit{S} has the symmetry (2) only if one
considers the contractions of indices between 
\textit{S} and \textit{T}. These cases reflect items (i)
and (ii) respectively of the following lemma.

\

\noindent \textbf{Lemma 2}: Let \textit{T} be a non-null
tensor of rank \textit{m} with \textit{2n} indices
contracted ($2n \leq m$). Let $\sigma$ be a permutation
such that 

\begin{equation}
T\,^{\sigma ({i_{1}}}\,^{{j_{1}}}\,^{{\cdots_{\ }
}}\,^{{i_{n}}}\,^{j_{n}}\,^{{f_1}}
\,^{{\cdots_{\ }
}}\,^{{f_{m-2n}})}
\end{equation}

\noindent describes an index configuration with 
\textit{m} free indices, such that after $j\,{_{p}}$
($p \leq n$) indices are lowered by the metric
terms as described in (3), one obtains the actual
index positions of \textit{T}. Suppose that (4)
does not vanish.
\textit{T} is zero if and only if at least one of the 
following items is satisfied.

\begin{itemize}

\item[(i)] Consider (4). Independently of any contraction,
there is a permutation $\rho$ acting on 
$i_1\,j_1\,\cdots\,i_n\,j_n$
such that
$$T^{\rho \circ \sigma (i_1\,j_1\,\cdots\,i_n\,j_n\,f_1\,
\cdots\,f_{m-2n})} = \epsilon \,
T^{\sigma (i_1\,j_1\,\cdots\,i_n\,j_n\,f_1\,
\cdots\,f_{m-2n})}$$
\noindent where $\epsilon=1$ or
$\epsilon=-1$
and at least one of the 
following items is satisfied.

\begin{itemize}
\item[(i.1)] $T^{\rho \circ \sigma (i_1\,j_1\,\cdots\,i_n\,j_n\,f_1\,
\cdots\,f_{m-2n})}$ is antisymmetric 
under one or more interchanges
of  $i\,{_{p}} j\,{_{p}}$ into
$j\,{_{p}} i\,{_{p}}$
for $p \leq n$

\item[(i.2)] $T^{\rho \circ \sigma (i_1\,j_1\,\cdots\,i_n\,j_n\,f_1\,
\cdots\,f_{m-2n})}$ is 
antisymmetric under one or more pair
interchange of 
$i\,{_{p}} j\,{_{p}}$
into
$i\,{_{q}} j\,{_{q}}$
for $p,q \leq n$.
\end{itemize}

\item[(ii)] There is an index character configuration such that 
\textit{T} is antisymmetric under a permutation $\pi$
acting on the contravariant indices (like (1))
and \textit{T} is invariant 
under the same permutation acting on
the corresponding covariant indices (like (2) 
with all indices in the same tensor).

\end{itemize} 
\noindent  \textbf{Proof:}  \cite{Pelavas}

\

For example, suppose that \textit{T} is a tensor of rank 6
with the symmetries

$$T^{\,i\,j\,k\,l\,m\,n} = - T^{\,k\,l\,i\,j\,m\,n},$$
$$T^{\,i\,j\,k\,l\,m\,n} = T^{\,j\,i\,k\,l\,m\,n},$$
$$T^{\,i\,j\,k\,l\,m\,n} = T^{\,i\,j\,l\,k\,m\,n},$$
$$T^{\,i\,j\,k\,l\,m\,n} = T^{\,i\,j\,k\,l\,n\,m}.$$

\noindent Consider the index configuration

\begin{equation}
T^{\,i\,k\,l}\,{_{\,i\,k\,l}}
\end{equation}

\noindent which is equivalent to zero. Due to the contraction
of the first and fourth indices, (5) is antisymmetric under
the interchange of $k^{+}$ and $l^{+}$, and is symmetric under 
the interchange of  $k^{-}$ and $l^{-}$. This is an 
example of item (ii) of lemma 2.

   There is one case not analyzed yet. A tensor \textit{T}
may vanish due to symmetries regardless of
any index contraction. This case has no practical
applications but the algorithm must recognize
what are the combinations of symmetries that cancel
the tensor. That recognition can be done at the moment of defining
the tensor, before the algorithm is executed.
From now on, suppose that tensors are not zero if there is
no index contraction.

Let $\mathcal{M}$ be
the set of all mathematically equivalent tensor products
generated  by all possible \textit{equivalent}
configurations of the indices of the tensors of 
$\mathcal{P}$ ($\mathcal{P}$ is a product of non-null
tensors when all indices are considered free). 
Definitions 4-6 can be extended to products 
of tensors by application on each factor. 

\

\noindent \textbf{Lemma 3}: Two elements of $\mathcal{M}$ have the 
\textit{index canonical configuration} with opposite signs
if and only if $\mathcal{P}$ is zero.

\noindent \textbf{Proof}: ($\Rightarrow$) If $\mathcal{P}$ is
mathematically equivalent to $\mathcal{P}$' and 
$-\mathcal{P}$' simultaneously then $\mathcal{P}=0$.

($\Leftarrow$) If $\mathcal{P}$ is zero, 
either one of the factors vanishes 
or, using lemma 1, there is a 
factor \textit{A} antisymmetric on
some of its indices (like (1)) such that 
$\mathcal{P}$ satisfies (2).
If one of the factors vanishes,  item (i) or
item (ii) of lemma 2 are obeyed. For all
possible cases, two equivalent opposite terms
with the \textit{index canonical configurations}
are generated due to the presence of contracted
antisymmetric indices.

\

In worst cases, the algorithm recognizes that a factor
is zero in step 7f, that
a product is zero in step 7h, and that a sum
is zero in step 10.

\

\noindent \textbf{Rule 1}: Consider a tensor product. The character of the
first (from left to right) index of a pair of summed indices (if some
exists) is contravariant, and the character of the second is
covariant. The same rule applies for the summed indices within a
single tensor.

\

\subsection{The algorithm}

 The algorithm is divided into
steps grouped by types of action that must be followed in increasing
order unless otherwise stated.  The main goal is to put the indices into a
canonical position. The canonical position involves the relative
position of the indices inside the tensor as well as their character.
Definition 6 is a precise specification of the canonical position.
All dummy indices are renamed and the original names are thrown away.
The renaming of the dummy indices takes into account their position
inside the tensors and the order of the tensors. Therefore the
renaming process leads to canonical names only after the dummy indices
have been put into a canonical position and an ordering for the
tensors has been established.

\

\noindent \textbf{Step 1}.(Expanding) Expand the tensor 
expression modulo the
coefficients. Select the tensor factors of the first term.\footnote{
For expressions consisting of  one tensor, only steps 6\textbf{
}and\textbf{ }9 are performed if the tensor has no symmetries; steps 7
and 9 are performed if the tensor has symmetries. Classes 
${S_{I}}$
and 
${S_{{\it II}}}$
are empty is this case.} From now on, consider how to put a
tensor product into a canonical form. Steps 1 to 9 are applied to all
terms of the expanded expression, one at a time.

\

\noindent \textbf{Step 2}.(Raising or lowering indices) If there are any
metric tensor with indices contracted with other tensors, the
corresponding indices are raised or lowered according to the
contraction character. The same goes for other special tensors built
from the metric.

\

\noindent \textbf{Step 3}.(Splitting in group I and II) Split the product in
two groups. Group I consists of tensors with symmetries. Group II
consists of tensors with no symmetries. Tensors of group I are placed
in the left positions and tensors of group II in the right positions
using the commutativity property of the product.

\

\noindent \textbf{Step 4a}.(Merging equal tensors) Tensors of group II with
the same name and same number of indices and with no free indices
merge into a new tensor. Suppose that each original tensor has 
$n$
indices and that there are 
$N$
equal tensors, then the new tensor has 
${\it nN}$
indices and  is totally symmetric under the interchange of the
groups of the  
$n$
indices.  After the merging, the resulting tensors are incorporated
into group I. The information about the relation with original tensors
is stored, since it will be used at the end of the algorithm to
substitute the new tensor with the original tensor names. 

\

\noindent \textbf{Step 4b}. Tensors of group I with same names, same
number of indices and same number of free indices merge to form a new
tensor. The new tensor has the same symmetry under the interchange of
group of indices as described in step 4a, and each group of indices
inherits the symmetries of the original tensors. In step 4a, only
tensors of group II with no free indices are merged. In this step, the
tensors may have free indices.\textbf{{\large  }}

\

\noindent \textbf{Step 5a}.(Sorting the tensor names) The tensors are
lexicographically sorted inside each group. Tensors with same name but
with different number of indices are sorted
according to the number of indices.

\

\noindent \textbf{Step 5b}. Tensors with the same name and same number of
indices are sorted according to the number of free indices. 

\

\noindent \textbf{Step 5c}. Tensors of group II with the same name, same
number of indices, and same number of free indices are sorted
according to the name of the 
first free index. Notice that the free indices have fixed
positions for tensors of group II.

\

\noindent \textbf{Step 6}.(Fixing the index character of group II) Consider
the tensors of group II. The summed indices that are contracted within
the tensors or those that are contracted with other tensors of group
II may have their character changed in order to obey rule 1. All other
summed indices (the ones contracted with tensors of group I) are put
covariant. The free indices remain untouched.\footnote{ This step can
be displaced and performed together with step 9}.

\

\noindent \textbf{Step 7a}.(Splitting in classes) Step 7 is performed for all
tensors of group I. Consider the first tensor of group I (the current
tensor). It has symmetries that can involve all indices or only some
of them.  The discussion that follows applies only for the indices
involved in the symmetries or contracted with indices involved in the
symmetries.  The pairs of summed indices not involved in the
symmetries must obey rule 1 and the free indices not involved in the
symmetries remain untouched throughout the whole algorithm.  Consider
the indices that can have their positions affected by the symmetry.
They are split into classes: 

Class 
$F^{+}$: contravariant free indices

Class 
${S_{{\it II}}}$: summed indices contracted with tensors of group II.

Classes  
$A^{+}$, 
$B^{+}$, 
$C^{+}$, 
$A^{-}$, 
$B^{-}$, 
$C^{-}$: summed indices inside the tensor

Class 
${S_{I}}$: summed indices contracted with tensors of group I.

Class 
$F^{-}$: covariant free indices

\noindent Let 
$f^{+}$, 
${s_{{\it II}}}$, 
$a$, 
$b^{+}$, 
$c^{+}$, 
${s_{I}}$, 
$F^{+}$, 
$A$, 
$B^{+}$, 
$C^{+}$, 
${S_{I}}$, 
$C^{-}$, 
$B^{-}$
and 
$F^{-}$
respectively. Classes  
$A^{+}$
and 
$A^{-}$
have the same size.  

       The summed indices inside the tensor are members of classes  
$A^{+}$, 
$B^{+}$, 
$C^{+}$, 
$A^{-}$, 
$B^{-}$
or 
$C^{-}$. 
When both indices of a pair of summed indices are involved in the
symmetries, the contravariant index is a member of class 
$A^{+}$, and the covariant is a member of class 
$A^{-}$. If a index of a pair of summed indices is involved with the
symmetries while the corresponding one is not, there are four cases.
Let us call 
$i^{+}$
the index involved in the symmetries and 
$i^{-}$
the corresponding index  not involved in the symmetries. These
indices are equal but have different characters  (here the signs + and
- do not describe the character). Suppose that the relative position
of  
$i^{+}$
 with respect to  
$i^{-}$
cannot be inverted due to the symmetries. If 
$i^{+}$
 is at the right of 
$i^{-}$, then  
$i^{+}$
is a member of  class 
$B^{+}$; if 
$i^{+}$
 is at the left, then  
$i^{+}$
is a member of class 
$B^{-}$. If the relative positions of  
$i^{+}$
 and  
$i^{-}$
can change, then 
$i^{+}$
is a member of  class 
$C^{+}$
or  
$C^{-}$
depending on whether 
$i^{+}$
is contravariant or covariant. 

        As soon as class A is determined, it is verified whether 
there are antisymmetric contracted indices. If so, 
the tensor is null and the algorithm returns to step 2 for the
next term.\footnote{The vanishing of the tensor
due to the presence of antisymmetric
contracted indices can be a natural consequence of the application
of steps 7d to 7f together with lemma 3. For the sake
of efficiency, it is better to verify the presence of antisymmetric
contracted indices as soon as class A is determined. }

        The indices of class  
${S_{I}}$
 are contracted with the indices of the tensors of group I. Some of
the latter indices may not be involved in the symmetries. They form
sub-class 
${S_{I}}^{0}$, which is further split into  
${S_{I}}^{0^{+}}$
and  
${S_{I}}^{0^{-}}$, 
corresponding to the indices of  
${S_{I}}^{0}$
contracted with the tensor to the right or to the left of the
current tensor respectively.  The remaining indices of 
${S_{I}}$
are split  into sub-classes. The indices of the current tensor
contracted with the next (to the right) tensor of group I form the
first sub-class (${S_{I}}\,^{1}$); 
the indices contracted with the next tensor form the second
sub-class (${S_{I}}\,^{2}$), 
and so on until the last tensor of group I. If the current tensor
is the first of group I, the sub-division in classes is complete;
otherwise the sub-division continues and the next sub-class consists
of the indices the current tensor contracted with the first tensor of
group I. The following sub-class consists of the indices contracted
with the second tensor of group I and so on until the last tensor of
group I that has not been considered yet.

\

\noindent \textbf{Step 7b}.(Fixing the index character of group I)  In order
to obey rule 1, the indices of class 
${S_{{\it II}}}$
are put contravariant. The corresponding indices of tensors of group
II have already been put covariant in step 6. The indices of class 
${S_{I}}$
are put contravariant if they are contracted with the tensor at the
right, or covariant if they are contracted with the tensor at the left
of the current tensor.  The indices of class 
$B^{+}$
are put contravariant and the corresponding ones are put covariant.
The indices of class 
$B^{-}$
are put covariant  and the corresponding ones are put contravariant.
The character of the indices of classes 
$F^{-}$
is maintained throughout the whole algorithm. The character of
classes  
$A^{+}$, 
$C^{+}$, 
$A^{-}$
and 
$C^{-}$
may still change.  

\

\noindent \textbf{Step 7c}.(Ordering and the numbering of the indices) This
step establishes the order of the indices of classes  
$F^{+}$, 
${S_{{\it II}}}$, 
$B^{+}$, 
$C^{+}$, 
${S_{I}}$,  
$C^{-}$, 
$B^{-}$
and 
$F^{-}$
 to be used in step 7e. Also, it specifies the numbers each index
receives in order for definition 3 to be applied. To begin with, let
us discuss the order of the indices of classes 
$F^{+}$
and 
$F^{-}$. First, sort the free indices of class  
$F^{+}$
using their original names. The first sorted free index is
substituted with the number 
$1$, the second by 
$2$, and so on until the 
$f^{+}$
th index, which is substituted with 
$f^{+}$. The same process is performed for the indices of class 
$F^{-}$, which receive the numbers 
$f^{+}\ ..\ a^{-} + 1$, $\cdots$, 
$f^{+}\ ..\ f^{-}$.\footnote{ The notation 
$f^{+}\ ..\ f^{-}$
means 
$f^{+} + {s_{{\it II}}} + a^{+} + b^{+} + c^{+} + {s_{I}} + c^{-}
 + b^{-} + a^{-} + f^{-}$. 
The variables 
$a^{+}$
and 
$a^{-}$
are equal to 
$a$.} 

      Class 
${S_{{\it II}}}$
consists of summed indices contracted with the tensors of group II.
At this point, the tensors of group II are ordered. The positions of
the indices of these tensors are used as a reference to order the
indices of class 
${S_{{\it II}}}$. 
The first summed index  in group II (from left to right) that is a
member of class 
${S_{{\it II}}}$
 is the first index of 
${S_{{\it II}}}$. 
It receives the number 
$f^{+} + 1$. 
The second receives the number 
$f^{+} + 2$
 and so on until  
$f^{+} + {s_{{\it II}}}$.  

          The indices of  classes  
$B^{+}$, 
$B^{-}$,  
$C^{+}$
and  
$C^{-}$
are contracted with indices that have no symmetries; therefore, they
have an ordering reference. They follow the same method used for the
indices of class 
${S_{{\it II}}}$. 

    Class 
${S_{I}}$
can have its sub-classes ordered. The first sub-class is 
${S_{I}}^{0^{+}}$, followed by sub-classes 
${S_{I}}\,^{1}$,  
${S_{I}}\,^{2}$
and so on, and the last sub-class is 
${S_{I}}^{0^{-}}$.  The indices of the classes  
${S_{I}}^{0^{+}}$
and  
${S_{I}}^{0^{-}}$
 can be ordered since they are contracted with indices not involved
in the symmetries. They follow the same method used for the indices of
class 
${S_{{\it II}}}$. 
The indices of the other sub-classes cannot be ordered at this
point. Also, the indices of class 
$A$
cannot be ordered at this point.

  Here follows, explicitly, the numbers reserved for each class:

Class 
$F^{+}$:  
$1$, 2, $\cdots$, 
$f^{+}$


Class 
${S_{{\it II}}}$:  
$f^{+} + 1$,  
$f^{+} + 2$, $\cdots$, 
$f^{+} + {s_{{\it II}}}$


Class 
$A^{+}$: 
$f^{+} + {s_{{\it II}}} + 1$, $\cdots$, 
$f^{+}\ ..\ a^{+}$

Class 
$B^{+}$: 
$f^{+}\ ..\ a^{+} + 1$, $\cdots$, 
$f^{+}\ ..\ b^{+}$

Class 
$C^{+}$: 
$f^{+}\ ..\ b^{+} + 1$, $\cdots$, 
$f^{+}\ ..\ c^{+}$

Class 
${S_{I}}$:  
$f^{+}\ ..\ c^{+} + 1$, $\cdots$, 
$f^{+}\ ..\ {s_{I}}$

Class 
$C^{-}$: 
$f^{+}\ ..\ {s_{I}} + 1$, $\cdots$, 
$f^{+}\ ..\ c^{-}$

Class 
$B^{-}$: 
$f^{+}\ ..\ c^{-} + 1$, $\cdots$, 
$f^{+}\ ..\ b^{-}$

Class 
$A^{-}$: 
$f^{+}\ ..\ b^{-} + 1$, $\cdots$, 
$f^{+}\ ..\ a^{-}$

Class 
$F^{-}$:  
$f^{+}\ ..\ a^{-} + 1$, $\cdots$, 
$f^{+}\ ..\ f^{-}$

\

\noindent \textbf{Step 7d}.(Generating the character configurations) Both
indices of a pair from class 
$A$
can change their positions due to the symmetries,  but the relative
position of some pairs may be fixed. The characters of the pairs, that
cannot invert the relative position, can be chosen such that they obey
rule 1. The characters of the remainder indices of class 
$A$
(let 
${a_{0}}$
be the number of pairs of class 
$A$
that can invert their relative position) and the characters of
indices of class 
$C$
cannot be chosen a priori.  Each pair has two states which are
contravariant-covariant and covariant-contravariant, corresponding to
the characters of the indices. The algorithm generates 
$2^{({a_{0}} + c)}$
possible character configurations by changing the states of pairs
that can invert their relative position.

  If the current tensor is totally symmetric, 
steps 7d and 7f need not be performed. The canonical form
can be obtained straightforwardly, avoiding the slowest steps.

\

\noindent \textbf{Step 7e}.(Applying the symmetries) At this point,  more
than one \textit{equivalent} configuration may have been generated.
The symmetries are applied to all configurations
in order to perform the following tasks.\footnote{ The application 
of the symmetries
is a straightforward procedure that can be implemented for each kind
of symmetry. In the case of the Maple system, tensors can be
represented by tables and the symmetries by indexing functions, which
can perform the tasks described in step 7e.} The contravariant indices are
pushed to the left positions as far as possible and the covariant
indices to right positions as far as possible. A sign change may be
generated if there are antisymmetric indices. The state configurations
that are members of the set of \textit{character normal
configurations} of the current tensor are selected. The symmetries are
applied to the selected configurations again in order to put classes  
$F^{+}$, 
${S_{{\it II}}}$, 
$A^{+}$, 
$B^{+}$, 
$C^{+}$, 
${S_{I}}$,  
$C^{-}$, 
$B^{-}$, 
$A^{-}$
and 
$F^{-}$
as closely as possible in the smallest position as prescribed in
definition 3a, and after that, the indices of classes  
$F^{+}$, 
${S_{{\it II}}}$, 
$B^{+}$, 
$C^{+}$,  
${S_{I}}\,^{0}$,
$C^{-}$, 
$B^{-}$, and 
$F^{-}$
 are put as closely as possible in their order as prescribed in
definition 3b (see step 7c for relabelling). The sub-set 
of  \textit{index normal configurations} is
selected. At this point, the
character and the positions of all classes have been determined.
Only the order of the indices of class 
$A$
and of the indices of the sub-classes of 
${S_{I}}$
(not including ${S_{I}}^{0}$) has not been
determined yet. 

\

\noindent \textbf{Step 7f}.(Generating \textit{equivalent} 
configurations of class $A$)  In general, the ordering of 
the indices of the sub-classes of ${S_{I}}$ depends on
the ordering of 
the indices of class $A$ and 
vice-versa. These classes must be ordered together.
For now, suppose that all classes  ${S_{I}\,^{i}}$ ($i>0$)
are empty. In this case, 
steps 7a to 7g can be performed for each factor
of the product since they are independent of each other.

   The aim of this step is
the generation of permutations of class $A$
that preserve the character arrangement of the indices.
The selected configurations of step 7e are
submitted to all possible re-orderings of class $A$ (class  
$A^{+}$
plus  
$A^{-}$)
allowed by \textit{induced symmetry} of the indices of
class $A^{+}$ together with the indices of class $A^{-}$.  
The \textit{character normal configurations} are selected. 

If the  \textit{induced symmetry} of class $A$ is totally
symmetric, the generation of permutation 
are not necessary since the
canonical positions can be obtained straightforwardly.

In many cases, it is sufficient to generate the re-orderings
allowed by the \textit{induced symmetries} of
class $A^{+}$
and  
class $A^{-}$ independently. This kind of re-ordering 
automatically maintain the character configuration.

\

\noindent \textbf{Step 7g}.(Selecting the \textit{index canonical
configuration}) The indices of all configurations are substituted by
their correspondent numbers (step 7c), and the \textit{index canonical
configuration} with respect to classes  
$F^{+}$, 
${S_{{\it II}}}$, 
$A^{+}$, 
$B^{+}$, 
$C^{+}$, 
${S_{I}}$, 
$C^{-}$, 
$B^{-}$, 
$A^{-}$
and 
$F^{-}$
is  selected. 

If two elements with the  \textit{index canonical
configuration} are selected and they have opposite signs 
then the current product is zero (lemma 3). In this case,
the algorithm returns to step 2 for the next term.

\

\noindent \textbf{Step 7h}.(Ordering indices of
the sub-classes of ${S_{I}}$) 
The order of the sub-classes of ${S_{I}}$
and the order of the indices of ${S_{I}}\,^{0}$ have already
been determined. The present step finds out the order of the indices
of sub-classes ${S_{I}\,^{i}}$ ($i>0$) of all tensors of
the current product.  
Consider class $A$ and all sub-classes  ${S_{I}\,^{i}}$ ($i>0$) of 
the first tensor. These classes together have a \textit{induced
symmetry}. The same can be stated about the other 
factors of the product. All indices of all classes $A$ and all
sub-classes  ${S_{I}\,^{i}}$ ($i>0$) of all factors are put
together  in order to
form a new tensor totally contracted. The
order of the indices in the new tensor 
follows the order of the factors and the order of appearance 
in each factor.  
The symmetries of the new
tensor is composed by all induced symmetries acting
in the corresponding indices (see example 2).    

By a recursive call of the algorithm, the indices of the new
tensor are in class 
$A$
and are ordered by the method described for this class. 
The dummy indices that come from sub-classes ${S_{I}\,^{i}}$
cannot invert their relative positions (within a pair)
hence step 7d need not generate character configurations 
for these indices.

If the \textit{induced symmetry} of the indices
of the sub-classes ${S_{I}\,^{i}}$
of a factor coincides with the actual symmetry of these 
indices (taking into account the contractions of class $A$)
then the indices of class $A$ of this factor
need not be included in the new tensor.

\

\noindent \textbf{Step 8}.(Recovering merged tensors) The symmetric (by group
of indices) tensors that have been formed by merging tensors with
equal names, equal number of indices and equal number of free indices
in step 4a and 4b are converted back by the inverse process to a
product of tensors with the original names but with the new positions
of indices
generated by the previous steps. 

\

\noindent \textbf{Step 9}.(Renaming the dummy indices)  At this stage of the
algorithm, the indices are in their final position. The dummy indices
are renamed following the rules: There are two cases. The first occurs
when the whole pair of dummy indices resides inside a tensor. If the
tensor name is 
$A$
and the number of indices is 
$m$, then the dummy index name will be 
${\it A\_m\_i\_j}$, where 
$i$
is the position of the contravariant index and 
$j$
is the position of the covariant index and 
$\_$
is some separator.\footnote{ The separator is a symbol not present in
tensor expressions.}  If the same name 
appears in other tensors of the product, no name
conflict is generated.  The second case occurs when the dummy indices
involve two tensors. Suppose that the first tensor has the name 
$A$
with 
$m$
indices and the second has the name 
$B$
with 
$n$
indices, then the dummy index will be renamed  
${\it A\_m\_i\_B\_n\_j}$
, where 
$i$
is the position of the contravariant index inside the tensor 
$A$
and 
$j$
is the position of the covariant index inside the tensor 
$B$. This renaming process proceeds from left to right. If a second
dummy index receives the same name, then the number 1 is appended in
its name  
${\it A\_m\_i\_B\_n\_j\_1}$, 
for example. If a third summed index receives the same name, then
the number 2 is appended in its name: 
${\it A\_m\_i\_B\_n\_j\_2}$, 
and so on.\footnote{ The method of appending a number to the
repeated dummy index names can be fully avoided if step 8 is performed
after step 9.}

\

\noindent \textbf{Step 10}.(Collecting equal tensor terms) After performing
steps 1 to 9 for all terms, collect equal tensor products and put the
coefficient factors into the canonical form. 

\

\noindent \textbf{Step 11}.(Sorting the addition)\footnote{ In general, this
step is not necessary when the algorithm is implemented
over multiple purpose computer algebra systems.} 
Each tensor product can be converted to a string by concatenating with
separators the names of the tensors and the indices in the order they
appear in the product. These strings are sorted. The order of the
terms in the sum is rearranged to be in the same order as the sorted
concatenated strings.

\

\subsection{Proof that the algorithm is a canonical function}

Let $\mathcal{T}$ be the set of all tensor expressions which obey
hypothesis 1. The algorithm 
described in section 2.2 is a function  
\mbox{$\mathcal{F} : \mathcal{T} \longmapsto \mathcal{T}$}.

\

\noindent \textbf{Theorem}: $\mathcal{F}$ is a canonical function.

\noindent \textbf{Demonstration}: The proof has three parts.

1. All operations performed by the algorithm obey the rules
of the tensor algebra, hence preserve the mathematical
equivalence of tensor expressions.

2. For all 
$\mathcal{E}\,_{1},\mathcal{E}\,_{2} \in \mathcal{T}$
such that
$\mathcal{E}\,_{1}=\mathcal{E}\,_{2}$,
$\mathcal{F}(\mathcal{E}\,_{1})\equiv\mathcal{F}(\mathcal{E}\,_{2})$.
After step 1, 
$\mathcal{E}\,_{1}$ and
$\mathcal{E}\,_{2}$
are sums of tensor products. Consider a generic tensor product.
It is clear 
from the sorting uniqueness that the position of
the tensors is unique after the
application of the algorithm. 
The indices of  tensors of group II and the indices of classes 
$F^{+}$, 
${S_{{\it II}}}$,  
$B^{+}$, 
${S_{I}}$,  
$B^{-}$
  and 
$F^{-}$
of tensors of group I go to a unique character configuration,
since this is a matter of convention.

    The proof that the indices of 
tensors of group I (including the merged tensors of
step 4) come to a unique configuration is consequence of
the fact that the algorithm runs over all index character
configurations of the indices of classes  
$A$
and 
$C$
and over  all allowed index position configurations of classes 
$A$
and 
${S_{I}}$. Only one configuration
is selected as the canonical configuration unless the 
product is zero (lemma 3). After step 9, the dummy indices have
canonical names, completing the canonicalization of a tensor
product. Steps 10 and 11 put a sum of tensor products
into the canonical form.

3. In item 2 is missing the details about the special
case when one tensor expression is zero, that is, if 
$\mathcal{E}=0$ then 
$\mathcal{F}(\mathcal{E})\equiv0$. From lemmas 1 
and 2 follow that the indices 
responsible for the 
cancelation of a factor are in class $A$ and 
for a product are in class $S_I$.
Since the algorithm generates the set of all 
mathematically equivalent tensor products
by permuting the indices of classes $A$ and $S_I$,
lemma 3 garantee that any null product inside the
tensor expression is recognized. Other 
non-null tensor
products that may still exist are put into a canonical 
form (item 2) guaranteeing the cancellation of a sum
of products in step 10.

\

In the algorithm, there are shortcuts for special kinds
of symmetries  avoiding the
generation of character configurations of step 7d 
or permutations of step 7f.
These shortcuts increase the efficiency 
but they must satisfy items 1-3 of the proof.

\

\subsection{Examples}

Example 1:  Consider the tensor expression 
$R\,^{i}\,{_{b}}\,{_{a}}\,{_{i}}$
(contraction of the Riemann tensor). Step 7a establishes that
classes 
$A$
and 
$F\,^{-}$
are the only non empty classes in this case. Class 
$A^{+}$
is  
$[i^{+}]$
, class 
$A^{-}$
is  
$[i^{-}]$
 and class  
$F\,^{-}$
is 
$[a, \,b]$. Step 7c establishes that the contravariant index 
$i$
receives the number 1, the covariant index 
$i$
receives  the number 2, and the indices 
$a$
and 
$b$
receive  the numbers 3 and 4 in this order, since 
$a$
precedes 
$b$. 
Step 7d establishes that there are two character configurations to
be considered:  
$R\,^{i}\,{_{b}}\,{_{a}}\,{_{i}}$
and  
$R\,{_{i}}\,{_{b}}\,{_{a}}\,^{i}$. 
The symmetries
are applied (step 7e) in order to put these configurations into the
equivalent ones   
$R\,^{i}\,{_{b}}\,{_{i}}\,{_{a}}$
 and  
$R\,^{i}\,{_{a}}\,{_{i}}\,{_{b}}$. 
Both are selected, since both are members of the set of 
\textit{character normal configurations}. No further changes are
generated by the symmetries, so the configurations that are members of
the set of \textit{index normal configurations} must be selected. The
indices are substituted  with their respective numbers, yielding
[1,4,2,3] and [1,3,2,4] respectively. Definition 3a with respect to
the partition  (1,1,2) -- corresponding to classes [1], [2] and
[3,4] -- selects both configurations. Definition 3b with respect to
class [3,4] selects [1,3,2,4] as the only member of the set of 
\textit{index normal configurations}. Since there are no
\textit{induced symmetries}, no more index configurations are
generated. The \textit{index canonical configuration} is  [1,3,2,4] . 
Therefore, the canonical form is  
$R\,^{i}\,{_{a}}\,{_{i}}\,{_{b}}$, where 
$i$
is 
${\it R\_4\_1\_3}$, as prescribed in step 9.

\

\noindent Example 2: Consider the tensor expression 
$T\,^{j}\,S\,^{k}\,^{l}\,T\,^{i}\,R\,{_{i}}\,{_{l}}\,{_{j}}\,{_{k}}$, 
where 
$S\,^{k}\,^{l}$
is a totally symmetric tensor and 
$R\,{_{i}}\,{_{l}}\,{_{j}}\,{_{k}}$
is the Riemann tensor. Step 3 splits the expression into group I: 
$[S\,^{k}\,^{l}, \,R\,{_{i}}\,{_{l}}\,{_{j}}\,{_{k}}]$
and group II:  
$[T\,^{j}, \,T\,^{i}]$. 
Step 4b merges 
$T\,^{j}$
and 
$T\,^{i}$
into one totally symmetric tensor (let be called 
${\it T\_T}\,^{j}\,^{i}$) 
and adds to group I. Group II is empty now. Step 5a establishes the
order for group I: 
$[R\,{_{i}}\,{_{l}}\,{_{j}}\,{_{k}}, \,S\,^{k}\,^{l}, \,{\it T\_T}\,^{j}\,^{i}]$. 
Step 7a establishes that the only non-empty class in this case is 
${S_{I}}$. 
For the first tensor one has 
${S_{I}}\,^{1}=[l, \,k]$
and 
${S_{I}}\,^{2}=[j, \,i]$. 
The order of these indices has not been established yet. These
are the only classes since all indices have been covered. 
Step 7b
fixes the character as:  
$[R\,^{i}\,^{l}\,^{j}\,^{k}, \,
S\,{_{k}}\,{_{l}}, \,{\it T\_T}\,{_{j}}\,{_{i}}]$. 
Step 7e converts 
$R\,^{i}\,^{l}\,^{j}\,^{k}$
into  
$R\,^{l}\,^{i}\,^{k}\,^{j}$ and maintains 
$S\,^{k}\,^{l}$
and
${\it T\_T}\,^{j}\,^{i}$ invariant. 
Step 7h generates a new tensor. Let us call 
$\_N\,^{l}\,^{i}\,^{k}\,^{j}\,{_{k}}\,{_{l}}\,{_{j}}\,{_{i}}$. 
It has the symmetries of the Riemann tensor (excluding the
cyclic symmetry) in the first four indices 
($l\,^{+}$,$i\,^{+}$,$k\,^{+}$,$j\,^{+}$);
is symmetric under the interchange of the fifth and sixth
indices ($k\,^{-}$,$l\,^{-}$) and 
is symmetric under the interchange of the seventh and eighth
indices ($j\,^{-}$,$i\,^{-}$). The algorithm is called
recursively, and  
$\_N\,^{l}\,^{i}\,^{k}\,^{j}\,{_{k}}\,{_{l}}\,{_{j}}\,{_{i}}$
is converted to 
$\_N\,^{l}\,^{i}\,^{k}\,^{j}\,{_{l}}\,{_{k}}\,{_{i}}\,{_{j}}$,
determining the order of the indices. 
After step 8 one has:  
$[R\,^{l}\,^{i}\,^{k}\,^{j}, \, 
S\,{_{l}}\,{_{k}}, \,T\,{_{i}}, \,T\,{_{j}}]$. 
Step 9 establishes that canonical form for the expression is  
$ R\,^{l}\,^{i}\,^{k}\,^{j}\,S\,{_{l}}\,{_{k}}\,T\,{_{i}}\,T\,{_{j}}$
where 
$l={\it R\_4\_1\_S\_2\_1}$,  
$i={\it R\_4\_2\_T\_1\_1}$,  
$k={\it R\_4\_3\_S\_2\_2}$
and   
$j={\it R\_4\_4\_T\_1\_1}$.

\

\section{An experimental implementation}

Algorithms to simplify tensor expressions have been implemented in
some computer al\-ge\-bra 
sys\-tems.\cite{Lee}\cite{Parker}\cite{Bogen}\cite{Stensor}\cite{Massoud} 
Some implementations use pattern
matching which requires a big database of tensor rules, and even
worse, sometimes the user must enter the rules. The
underlying method used by the Ricci package\cite{Lee} seems similar 
in some aspects to the
method presented here. All these implementations have not solved the
dummy index problem, therefore the main simplifier spends a long time
or cannot simplify tensor expressions with many dummy indices. 

In this section, I present an experimental implementation of the
algorithm described in section 2 over the Riemann 
package.\cite{Portugal} The new package can be obtained from web
sites,\footnote{ See the addresses:
http://www.cbpf.br/\symbol{126}portugal/Riegeom.html or
http://www.astro.queensu.ca/\symbol{126}portugal/Rie\-geom.html}
and my purpose is to supersede the Riemann package in the near future
with the new functionalities introduced to deal with tensor components
abstractly. 

The function \textit{normalform} uses the algorithm to put tensor
expressions into normal forms. No attempt has been made to put the
output into the canonical form, that is, ordered with respect to the
sum and to the product of terms, since this last step is unnecessary
for the purpose of any general computer algebra system.

In the examples below, contravariant indices have positive signs and
covariant indices  have negative signs. Tensors are indexed variables
and their symmetries are declared by the command
\textit{definetensor}. Some tensors are pre-defined: Christoffel
symbols, Riemann tensor, Ricci tensor and Ricci scalar are pre-defined
with the names 
$\Gamma$
(Christoffel symbols) and 
$R$
(Riemann, Ricci and Ricci scalar). More details can be found in the
new help pages for the commands \textit{normalform},
\textit{definetensor} and \textit{symmetrize}, and in the help pages
of the Riemann package. 

Expressions that are zero due to tensor symmetries are readily
simplified:\footnote{ All calculations have been performed in a Pentium
120 MHz with 32 Mb of RAM, running Maple V release 5 over Windows
95.}

\mbox{{\tt >}} readlib(showtime)():%

\mbox{{\tt >}} definetensor(T[i,j,k,l], sym[1,2] and asym[3,4]);%

\[
T\,^{i}\,^{j}\,^{k}\,^{l}
\]

time = 0.01, bytes = 13478

\mbox{{\tt >}} expr1 :=
printtensor(R[i,j,k,l]*T[-i,-k,-j,-l]);

\[
R\,^{i}\,^{j}\,^{k}\,^{l}\,T\,{_{i}}\,{_{k}}\,{_{j}}\,{_{l}}
\]

time = 0.05, bytes = 8124

\mbox{{\tt >}} normalform(expr1);

\[
0
\]

time = 0.12, bytes = 90619

\mbox{{\tt >}} expr2 :=
printtensor(T[i,j,k,l]*V[-i]*V[-j]+V[b]*V[a]*T[-a,-b,l,k]);

\[
T\,^{i}\,^{j}\,^{k}\,^{l}\,V\,{_{i}}\,V\,{_{j}} + V\,^{b}\,V\,^{a
}\,T\,{_{a}}\,{_{b}}\,^{l}\,^{k}
\]

time = 0.01, bytes = 8042

\mbox{{\tt >}} normalform(expr2);

\[
0
\]

time = 0.30, bytes = 338050

\mbox{{\tt >}} off;

Polynomials constructed with the Riemann tensor exemplify the
algorithm's worst performance. No special techniques have been
implemented for the kind of symmetries of the Riemann tensor. In the
next example, all possible ways to write the scalars formed by the
product of two Riemann tensors are generated. The  40.320 expressions
reduce to 4 independent non-null forms if the cyclic identity of the
Riemann tensor is not considered. In the next section, one can verify
that the cyclic identity reduces to 3 independent scalars (cf. ref.
\cite{Fulling}).\footnote{ Simplified names for the dummy indices
are used for this demonstration. This choice does not provide
truly canonical names.} 

\mbox{{\tt >}} S :=
\{op(combinat[permute]([a,b,c,d,-a,-b,-c,-d]))\}:

\mbox{{\tt >}} nops(S);

\[
40320
\]

\mbox{{\tt >}} readlib(showtime)():

\mbox{{\tt >}} S1 :=
map(x-\mbox{{\tt >}}abs(normalform(R[op(1..4,x)]*R[op(5..8,x)])), S);

\begin{eqnarray*}
\lefteqn{\{0, \, \left|  \! \,R\,^{{\it R1R1}}\,^{{\it R2R2}}\,^{
{\it R3R3}}\,^{{\it R4R4}}\,R\,{_{{\it R1R1}}}\,{_{{\it R2R2}}}\,
{_{{\it R3R3}}}\,{_{{\it R4R4}}}\, \!  \right| , \, \left|  \! \,
R\,^{{\it R1R1}}\,^{{\it R2R3}}\,^{{\it R3R2}}\,^{{\it R4R4}}\,R
\,{_{{\it R1R1}}}\,{_{{\it R3R2}}}\,{_{{\it R2R3}}}\,{_{{\it R4R4
}}}\, \!  \right| , } \\
 & &  \left|  \! \,R\,{_{{\it R1R1}}}\,{_{{\it R2R2}}}\,R\,^{
{\it R1R1}}\,^{{\it R2R2}}\, \!  \right| , \, \left|  \! \,(\,
{\rm }R)\, \!  \right| ^{2}\}\mbox{\hspace{262pt}}
\end{eqnarray*}

time = 35094.97, bytes = 21215820858

\noindent The program spends less than one second per expression on average.
Next, one example involving the product of three Riemann tensors is
provided:

\mbox{{\tt >}} normalform(R[-c,-d,m,n]*R[a,b,d,c]*R[-n,-m,-b
,-a]);

\[
 - R\,^{{\it R1R3}}\,^{{\it R2R4}}\,^{{\it R3R1}}\,^{{\it R4R2}}
\,R\,^{{\it R1R3\_1}}\,^{{\it R2R4\_1}}\,{_{{\it R1R3}}}\,{_{
{\it R2R4}}}\,R\,{_{{\it R3R1}}}\,{_{{\it R4R2}}}\,{_{{\it 
R1R3\_1}}}\,{_{{\it R2R4\_1}}}
\]

time = 26.41, bytes = 6303158

\mbox{{\tt >}} off;

\section{Simplification of tensor expressions}

The algorithm of section 2 achieves a full simplified form of the
tensor expressions if the tensors do not obey side identities. To
accomplish the simplification of tensor expressions obeying side
relations, the Gr\"obner basis method is used. The general strategy
is to put the tensor expression and the side relations into a
canonical form of the algorithm of section 2, and simplify the new
tensor expression with respect to the new side relation.

Here follows an example of how the  Gr\"obner basis method is used
to simplify the expression 
$R\,^{a}\,^{b}\,^{c}\,^{d}\,R\,{_{a}}\,{_{c}}\,{_{b}}\,{_{d}} - 
{\it 1/2}\,R\,^{a}\,^{b}\,^{c}\,^{d}\,R\,{_{a}}\,{_{b}}\,{_{c}}\,
{_{d}}$
.

\mbox{{\tt >}} expr :=
printtensor(R[a,b,c,d]*R[-a,-c,-b,-d]-1/2*R[a,b,c,d]*R[-a,-b,-c,-d]);

\[
{\it expr} := R\,^{a}\,^{b}\,^{c}\,^{d}\,R\,{_{a}}\,{_{c}}\,{_{b}
}\,{_{d}} - {\displaystyle \frac {1}{2}} \,R\,^{a}\,^{b}\,^{c}\,
^{d}\,R\,{_{a}}\,{_{b}}\,{_{c}}\,{_{d}}
\]

\mbox{{\tt >}} EXPR := normalform(expr);

\[
{\it EXPR} := R\,^{{\it R1R1}}\,^{{\it R2R3}}\,^{{\it R3R2}}\,^{
{\it R4R4}}\,R\,{_{{\it R1R1}}}\,{_{{\it R3R2}}}\,{_{{\it R2R3}}}
\,{_{{\it R4R4}}} - {\displaystyle \frac {1}{2}} \,R\,^{{\it R1R1
}}\,^{{\it R2R2}}\,^{{\it R3R3}}\,^{{\it R4R4}}\,R\,{_{{\it R1R1}
}}\,{_{{\it R2R2}}}\,{_{{\it R3R3}}}\,{_{{\it R4R4}}}
\]

\mbox{{\tt >}} side\_rel :=
printtensor(R[a,b,c,d]*symmetrize(R[-a,-b,-c,-d],cyclic[b,c,d]));

\[
{\it side\_rel} := R\,^{a}\,^{b}\,^{c}\,^{d}\,({\displaystyle 
\frac {1}{3}} \,R\,{_{a}}\,{_{b}}\,{_{c}}\,{_{d}} + 
{\displaystyle \frac {1}{3}} \,R\,{_{a}}\,{_{c}}\,{_{d}}\,{_{b}}
 + {\displaystyle \frac {1}{3}} \,R\,{_{a}}\,{_{d}}\,{_{b}}\,{_{c
}})
\]

\mbox{{\tt >}} SR := normalform(side\_rel);

\[
{\it SR} := {\displaystyle \frac {1}{3}} \,R\,^{{\it R1R1}}\,^{
{\it R2R2}}\,^{{\it R3R3}}\,^{{\it R4R4}}\,R\,{_{{\it R1R1}}}\,{
_{{\it R2R2}}}\,{_{{\it R3R3}}}\,{_{{\it R4R4}}} - 
{\displaystyle \frac {2}{3}} \,R\,^{{\it R1R1}}\,^{{\it R2R3}}\,
^{{\it R3R2}}\,^{{\it R4R4}}\,R\,{_{{\it R1R1}}}\,{_{{\it R3R2}}}
\,{_{{\it R2R3}}}\,{_{{\it R4R4}}}
\]

\mbox{{\tt >}} simplify(EXPR,\{SR=0\});

\[
0
\]

  From this example one notices that the scalars 
$R\,^{a}\,^{b}\,^{c}\,^{d}\,R\,{_{a}}\,{_{c}}\,{_{b}}\,{_{d}}$
and 
$R\,^{a}\,^{b}\,^{c}\,^{d}\,R\,{_{a}}\,{_{b}}\,{_{c}}\,{_{d}}$
are not independent.

\section{Conclusion}

An algorithm to simplify tensor 
expressions based on computable definitions
have been described. It has two parts. 
In the first part, the expression is
put into a canonical form, taking into account symmetries with respect
to index permutations and the renaming of dummy indices. The
definition of the canonical form involves some conventions that can be
changed without disqualifying the definition. The conventions are
based on the implementation simplicity and on a readable display for
tensor expression. In the second part, cyclic identities or more
general kinds of tensor identities are addressed through the
Gr\"obner basis method. The expression and the side relations are
both put into a canonical form for the Gr\"obner method to work
successfully. 

In this work, a precise definition of the canonical form
for tensor expressions is provided. 
No restriction is imposed on the kind of symmetries that the tensors
can obey. For most of the symmetries that occur in practise, the
algorithm is very fast. The symmetries of the Riemann tensor reveal
the algorithm's worst performance, but even in in this case it is
useful for practical applications. 

The invariant renaming of the dummy indices plays an important role in
the efficiency of the algorithm, since it neutralizes the symmetries
that come from dummy index renaming. This is a solution for the dummy
indices problem mentioned in the introduction.

An experimental implementation of the algorithm over the Riemann
package\cite{Portugal} is available free from web sites.\footnote{
See footnote 12.} All calculations and timings presented in this
work can be reproduced in the Maple system. 

\

\noindent \textit{\textbf{Acknowledgements:}}

I thank Ray McLenaghan and Keith Geddes for pointing out
references and for the kind hospitality at the University of Waterloo.
This work was made possible by a fellowship from
CAPES, Minist\'erio da Educa\c{c}\~ao e do
Desporto, Brazil.

\

\

\

\end{document}